# Enhancing Human-Robot Collaboration through Existing Guidelines: A Case Study Approach

Yutaka Matsubara[1[0000-0002-1111-0761]], Akihisa Morikawa[2[0000-0005-1579-1479]],

Daichi Mizuguchi[3[0009-0005-2020-948X]] and Kiyoshi Fujiwara[4[0000-0003-2535-7922]]

[1] Nagoya University, Nagoya, Japan
`yutaka@ertl.jp`
[2] IMAGINARY Corporation, Nagoya, Japan
`morikawa@imaginary-inc.jp`
[3] Atelier Corporation, Tokyo, Japan
`daichi.mizuguchi@ateliet-inc.com`
[4] National Institute of Advanced Industrial Science and Technology (AIST), Tsukuba, Japan
`k-fujiwara@aist.go.jp`

**Abstract.** As AI systems become more prevalent, concerns about their development, operation, and societal impact intensify. Establishing ethical, social, and safety standards amidst evolving AI capabilities poses significant challenges. Global initiatives are underway to establish guidelines for AI system development and operation. With the increasing use of collaborative human-AI task execution, it's vital to continuously adapt AI systems to meet user and environmental needs. Failure to synchronize AI evolution with changes in users and the environment could result in ethical and safety issues. This paper evaluates the applicability of existing guidelines in human-robot collaborative systems, assesses their effectiveness, and discusses limitations. Through a case study, we examine whether our target system meets requirements outlined in existing guidelines and propose improvements to enhance human-robot interactions. Our contributions provide insights into interpreting and applying guidelines, offer concrete examples of system enhancement, and highlight their applicability and limitations. We believe these contributions will stimulate discussions and influence system assurance and certification in future AI-infused critical systems.

## 1 Introduction

With the rapid advancement of AI, there has been a proliferation of AI systems. However, predicting the full impact of AI on humans and society during the development and operation of these systems is challenging. Moreover, the continuous evolution of AI through retraining and algorithm improvements introduces uncertainties during operation. Ethical [1-3], social concerns [4], and safety [5] are vigorously debated worldwide. International standards for developing and managing AI systems are being established to realize AI systems that meet human and societal needs [6-9].

Alongside the increasing prevalence of AI systems, there is a growing trend of human-AI collaboration, not only in tasks where AI replaces humans but also in scenarios where humans and AI-infused robots cooperate. In such collaborations, it's desirable for AI systems to adapt to users and environments continuously to fulfill their objectives and requirements. Ensuring ethics and safety involves understanding and accommodating the evolution of AI systems by users and environments. Failure to adapt to these changes may lead to ethical and safety issues, highlighting the necessity of mutual adaptation between users, environments, and AI systems. Although discussions on human-AI collaboration are ongoing, standardization efforts are still nascent (e.g. human-machine teaming is discussed and standardized as ISO/IEC PWI 42109).

Guidelines have been developed to facilitate human-robot interaction, emphasizing requirements for fostering collaboration and interaction. However, existing guidelines lack examples of system development, analyses and assurances incorporating human-AI collaboration.

This paper aims to evaluate the effectiveness and suitability of existing guidelines [10-12] through a case study of human-robot collaboration systems. We developed a virtual human-robot collaboration system and its system model using the Functional Resonance Analysis Method (FRAM) [13] to assess whether the system aligns with the requirements and evaluation metrics specified in these guidelines. Additionally, we discuss the contributions of existing guidelines to system development and identify their limitations to provide insights into their efficacy. Based on these findings, future discussions will concentrate on enhancing our co-evolution guidebook [14] with additional requirements.

The main contributions of this paper are:

- Demonstrating methods to interpret existing guidelines to apply their requirements and evaluation metrics to humans-AI collaboration systems.
- Providing concrete examples of system improvements based on whether they meet the specified requirements or not.
- Evaluating the applicability and limitations of existing guidelines concerning system evaluation, thereby guiding future discussions in the development of human-robot collaboration systems.

The paper follows this structure: Section 2 introduces the target system, a virtual human-robot collaboration. Section 3 utilizes FRAM to model the system, emphasizing human and robot functionalities and their interrelations. In Section 4, we present a use case to evaluate the model using existing guidelines. Section 5 discusses the applicability and limitations of these guidelines. Finally, we conclude and outline future directions.

## 2      Target System

We have developed a specific system for evaluating human-robot collaborative work in virtual environment. The system's objective is to assemble multiple components and produce finished products while ensuring worker safety and continuously improving efficiency during operation.

Fig. 1 illustrates the overview of the target system. The system involves multiple workers and automated mobile robots (AMRs) collaborating within the same workspace. Stations within the workspace feature spaces for holding components and workbenches. Workers produce products at the workbenches, with each station dedicated to producing different products requiring various components. Some products may also utilize components from other products.

The primary responsibility of the AMRs is to minimize wait times for required components by pre-positioning them to prevent shortages at each station. AMRs follow predetermined transport routes and halt transport temporarily if there's a risk of collision with surrounding objects (workers, other AMRs, field equipment, obstacles). Additionally, robots recognize workers and adjust their actions based on individual skills, such as slowing supply rates for less experienced  workers or maintaining longer safety margins.

Workers are primarily responsible for sending the specified number of final products out of the workspace. They repeatedly process components at each station to produce final products and move between stations, leveraging personal perception to gather information on robots, other workers, obstacles, and component locations. They adapt their actions flexibly based on updated AMR's behaviors.

The control center oversees the entire system, monitoring the status of all robots, workers, and components, recalculating the shortest transport routes as situations change, and issuing instructions accordingly. For instance, if workers become more skilled and reduce processing times, the control center adjusts plans to increase robot supply speeds, improving overall productivity. Moreover, if a robot malfunctions and can't transport components, workers take over the role. Workers' perceptions and robot sensors gather information on station positions, obstacles, workers, and component types and locations within the workspace, although recognition may be hindered by physical obstructions or distance. By delineating the roles and responsibilities of humans and AMRs in the collaborative system, this evaluation aims to provide insights into its effectiveness and potential improvements.

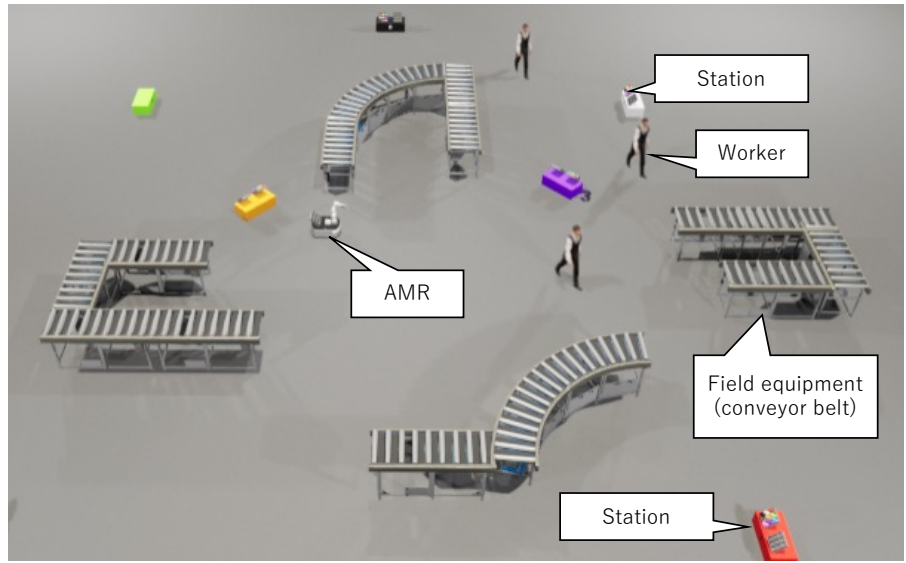

**Fig. 1.** Overview of the target system running in virtual environment.

## 3      System Modeling

### 3.1      Functional Resonance Analysis Method (FRAM)

To evaluate the interaction between humans and machines in the system proposed earlier, we adopted FRAM, which visually represents human-machine interactions. FRAM, developed by E. Hollnagel, is a modeling method for resilience systems. It focuses on functions and variability, where functions depict system operations and their interconnections, and variability indicates the variability in function outputs. Unlike linear causality or cause-effect reasoning, FRAM suggests that when the variability of two or more functions coincides, they can either attenuate or amplify each other, akin to resonance phenomena. Thus, one function's variability can influence the variability of other functions.

In [15], the system is described in terms of "functions" and the "interactions/correlations" between them. These interactions/correlations are classified into six elements:

- Input (I): Triggers the activation of a function to produce results described in Output.
- Output (O): Outputs the results of a function's execution (material, energy, information).
- Time (T): Temporal conditions affecting how a function is executed (constraints, resources).
- Control (C): Control conditions monitoring and governing a function.
- Preconditions (P): Preconditions that must be satisfied before a function is executed.

- Resource (R): Resource conditions required for a function's execution (material, energy, capabilities, human resources).

## 3.2    System Model for the Target System

For the initial FRAM model, we focused on the following aspects:

- We envisioned a system where humans and machines engage in mutual learning and collaborative work, with each influencing the other's behavior and improving over time. Hence, humans and machines assume equal roles. To clarify this, the model is divided into distinct human and machine parts, mirroring each other's structures.
- Accordingly, the internal modeling of humans is not based on actual human thought processes or responses.
- We described the model based on the system defined in the previous chapter while maintaining a certain level of abstraction.

The behavior of AMRs and workers is as follows:

- AMRs estimate self-position through sensing, gather information on other AMRs, workers, obstacles, and component arrangements, predict the behavior of others based on acquired information and past history, and execute movement, component retrieval, or placement based on goals and predictions while avoiding collisions during movement.
- Workers estimate their position through perception and gather information on AMRs, other workers, obstacles, and component arrangements. They predict the behavior of others based on acquired information and past experiences. Actions such as movement, component retrieval, placement, or processing are then executed based on these predictions and goals. All acquired information are saved for future use.
- AMRs and workers mutually learn and predict each other's behaviors, which are expected to improve over time based on past interactions.

The initial FRAM model is depicted in Fig. 2. The top half shows various functions of AMRs represented by red hexagonal nodes, while the bottom half represents worker behavior described as functions in green nodes. Each function is grouped according to its role: recognition, prediction and learning, judgment, and action, aligned from left to right, with the rightmost white node represents the reflection of real-world component operation results. The mutual relationship between AMRs and workers is limited to their respective movements and perceptions (lines intersecting at the left of the center) and component operation result perceptions. The gray nodes positioned at the top, bottom, and left edges represent external inputs, indicating factors like time constraints and work objectives.

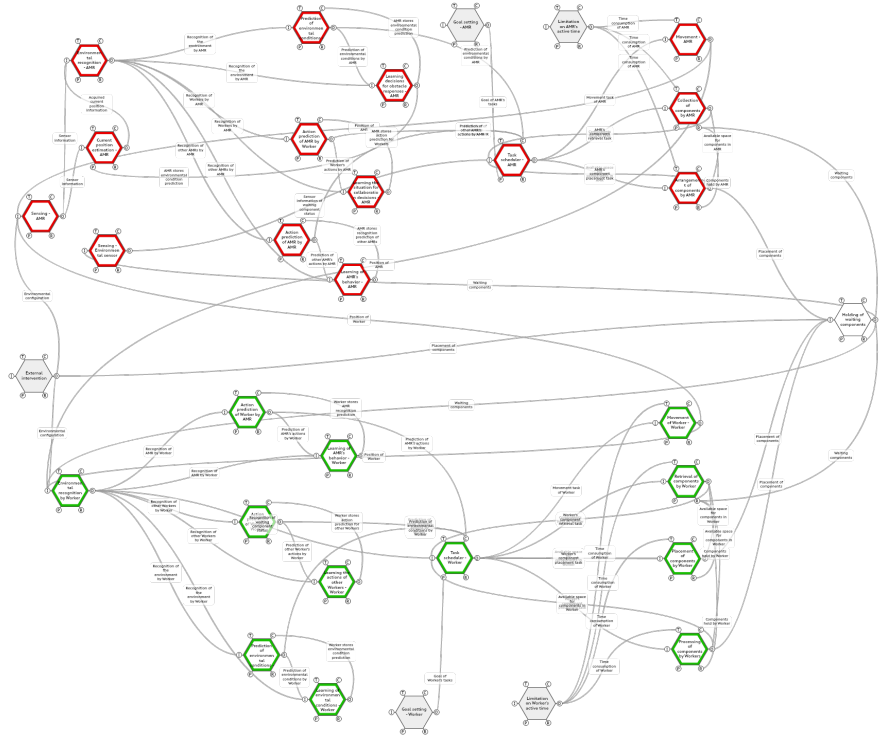

**Fig. 2.** Initial FRAM Model for the target system

## 4      Enhancing Human-Robot Collaboration System

### 4.1      Requirements and evaluation metrics

To evaluate and improve the effectiveness of human-robot collaboration and interaction in the target system, we extracted 16 requirements and evaluation metrics from the previous studies [10-12].

While 18 requirements are systematically organized as design guidelines applicable to human-AI interaction, distilled from a survey of over 150 AI-related design recommendations based on different interactions stages in [10], we focused on seven requirements related to collaboration and interaction between humans and robots under the stage category of "Over time" (G12-G18). In [11], four key challenges that need to be addressed from a human-robot teaming perspective are proposed. In [12], 45 papers on human-robot collaboration are categorized based on their claimed achievements, extracting five representative results.

**Table 1.** Evaluation metrics and results with existing guidelines and studies.

| Existing Guidelines or studies | Evaluation Metrics | Results | | | |
|---|---|---|---|---|---|
| | | C | AE | AN | N/A |
| | Remember recent interactions | | ✓ | | |
| | Learn from user behavior | ✓ | | | |
| Guidelines for Human-AI Interaction [10] | Update and adapt cautiously | | ✓ | | |
| | Encourage granular feedback | | ✓ | | |
| | Convey the consequences of user actions | | ✓ | | |
| | Provide global controls | | ✓ | | |
| | Notify users about changes | | ✓ | | |
| Shared Control in Human Robot Teaming [11] | Novel situations | | ✓ | | |
| | Context-aware communication | | ✓ | | |
| | Performance vs preference | | ✓ | | |
| | Interruptions and cognitive burden | | ✓ | | |
| Human-Robot Collaboration and Machine Learning [12] | Precision of movement | | | ✓ | |
| | Robustness | | ✓ | | |
| | Proof of concept | | | | ✓ |
| | Performance improvement | ✓ | | | |
| | Reduction of physical workload | | | ✓ | |

The results of evaluating these metrics for the FRAM model depicted in Fig. 2 are shown in Table 1. In this results, 'C' means that we confirmed the target system meet the requirement or metrics, 'AE' means that the requirement is applicable and employed to the target system, 'AN' means that the requirement is applicable, but not employed, and 'N/A' means the requirement is not applicable.

Here, the metric 'Proof of Concept' is marked as 'N/A' because it does not serve as a request for model improvement. Additionally, we consider the two items marked 'AN' to be strongly dependent on implementation and believe they should be revisited in the future.

## 4.2     Consideration of Improvement Measures

Table 2 presents the results of partial evaluation and requirements to improve the metrics. For example, "Update and adapt cautiously" serves as a metric to prevent disruptive system updates. In our system, workers and AMRs learn from each other and update their behaviors accordingly. However, there was no mechanism in place to prevent updates that could harm workers. This evaluation metric helped us recognize the need to add two functionalities: "Evaluation of differences between predictions and actual measurements" and "Authorizing learning function updates."

Building on the above examples and similar analysis activities, we identified the following eleven additional functionalities to further enhance human-robot collaboration systems. These new functionalities are designed to meet the metrics partially or in combination.:

- Functionalities related to prediction for workers, AMRs, and the environment:
  - Recording of predictions
  - Recording of actual measurements
  - Evaluation of differences between predictions and actual measurements
  - Updating prediction functions.
  - Pre-evaluating of learned prediction function updates.
- Functionalities related to communication between humans and robots:
  - Generating and sending messages.
  - Receiving and interpreting messages.
  - Preference requesting.
- Functionalities for robot updates:
  - Configuring learning function settings.
- Monitoring functionalities:
  - Monitoring task progress.
  - Suppressing activity.

Considering the identified improvement measures, we enhanced the FRAM model and the system accordingly. As a result, the FRAM model was improved to reflect the modifications. Figure 3 depicts the updated FRAM model, with newly added function nodes represented in slightly different colors.

In the process of implementing these enhancements on the FRAM model, we also made some adjustments to the connections with related peripheral functions. For instance, for two functionalities based on G14, "Pre-evaluating learned prediction functions" and "Updating prediction functions," we reconfigured the relationship between prediction and learning functions. The improved prediction model undergoes evaluation through the newly added "Evaluation of next-generation predictors" function, and if appropriate, the "Update prediction model" function updates the prediction function accordingly.

Since controlling the timing of learning reflection and adjusting learning parameters is not straightforward for humans, these functions are only specified for AMRs in the model.

**Table 2.** Evaluation results (Partial)

| Metrics | Explanations | Analytical results | Derived measures |
|---|---|---|---|
| Update and adapt cautiously (G14) | Limit disruptive changes when updating and adapting the AI system's behaviors. | The initial FRAM model employs continuous learning and predictive decision-making. However, it's noted that changes in behavior due to learning should be accompanied by appropriate evaluation, particularly regarding safety. | Implemented a process for pre-evaluating and authorizing prediction function updates.<br>• Updating prediction functions.<br>• Pre-evaluating of learned prediction function updates. |
| Encourage granular feed-back(G15) | Enable the user to provide feedback indicating their preferences during regular interaction with the AI system. | This aspect is currently not addressed in the initial FRAM model. However, it is essential for ensuring steady interaction between workers and AMR, forming feedback loops for interactive decision-making and task adaptation. | Added direct feedback mechanisms that reflect results in operations, enhancing real-time decision-making and task adaptability. <br> • Generating and sending messages. <br> • Receiving and interpreting messages. <br> • Monitoring task progress. <br> • Suppressing activity. |
| Novel situations (key challenge 1) | Robots must adapt to untrained tasks and environments, requiring robust risk management. | This aspect is not currently addressed in the initial FRAM model. While it corresponds to the predictive abilities of the AM-R system, it lacks specific specifications for handling scenarios not present in the behavior prediction history of workers. | Evaluate predictive accuracy based on recorded data and reflect this in actions to effectively manage risks in novel situations. <br> • Recording of prediction <br> • Recording of actual measurements <br> • Evaluation of differences between predictions and actual measurements <br> • Suppressing activity |
| Performance vs preference (key challenge 3) | The approach that delivers optimal performance may not always align with human preferences. | The current FRAM model incorporates performance as a metric; however, it fails to adequately recognize and integrate worker preferences. | Introduced functionality to input user preferences, aiding their integration into operational strategies. <br> • Generating and sending messages. <br> • Receiving and interpreting messages. <br> • Preference requesting |

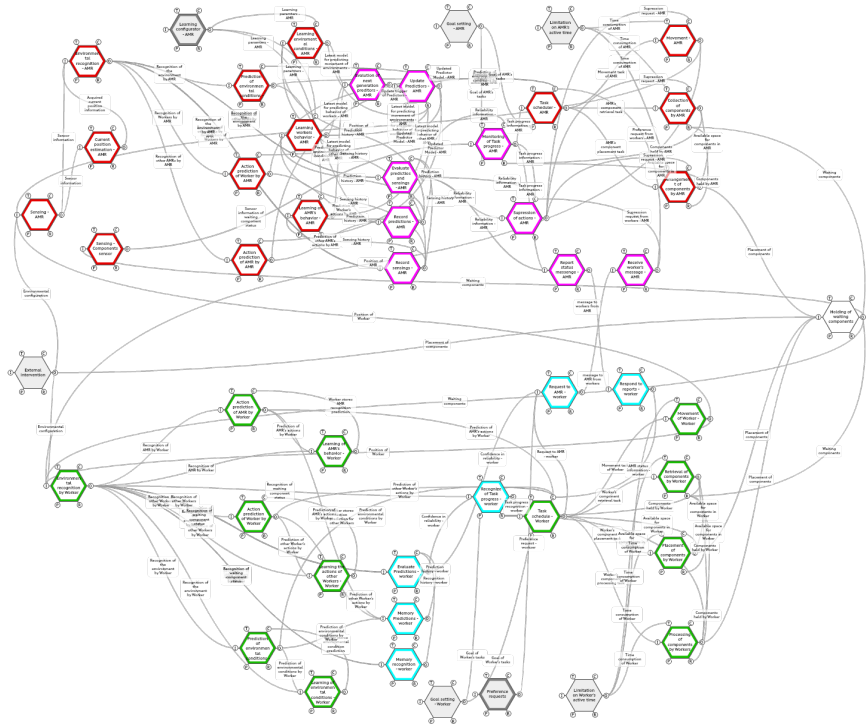

**Fig. 3.** Improved FRAM Model for the target system

## 5    Discussion

We explored enhancement strategies and mechanisms informed by design guidelines for human-AI interaction and key challenges in human-robot teaming [10-12]. We identified functional requirements, particularly from [10, 11], to foster mutual understanding between humans (workers) and robots (AMRs) and enable post-verification. Specifically, we implemented the following functionalities:

- Notifying each other of situational changes.
- Recording history of predictions and actual actions.
- Facilitating functional changes (updates).

Enabling mutual understanding between humans and robots allows for sharing interruptions or changes in tasks, potentially reducing the psychological burden on humans. Continuous adaptation of humans, robots, and the environment is crucial for reducing physical and psychological burdens while improving task efficiency through human-robot collaboration.

We encountered several challenges in evaluating using existing guidelines. Especially in utilizing [12], more detailed considerations are needed to derive concrete requirements for evaluation metrics marked as 'N/A' and 'AN' due to their abstraction. This implies that these metrics should be introduced at the initial design phase, and the target systems should be designed and implemented to meet the requirements according to these metrics.

While promoting mutual understanding between humans and robots is desirable, evaluating its contribution to the fundamental workplace objectives (e.g., maintaining/improving safety, work efficiency, user-friendliness) is challenging and not explicitly defined as a requirement for lifecycle management. The evaluation results also indicate a gap between system-level requirements and social level goals like ethics and well-being [16,17]. Unfortunately, the existing guidelines for human-AI or human-robot interactions can only partially fill these gaps. To address this issue, we need to discuss how system-level requirements, including AI system functionalities and human satisfaction and ethics, can be derived from social level goals. Through this discussion, we will specify requirements for logical environments, including policies, rules, and stakeholders (policy and/or rule makers, assessors of social level goals).

Furthermore, the methodology for argumentation is not adequately addressed in the existing guidelines. We advocate for a framework that encompasses comprehensive requirements for all stakeholders, including developers, users, and assessors. Continuous evaluation and systematic argumentation within this framework are crucial to ensure that the evolving needs of all stakeholders are effectively addressed and that both social and system-level goals are consistently met.

## 6    Conclusion

In this paper, we evaluated existing guidelines' applicability in human-robot collaborative systems, identify their effectiveness, and discuss limitations. Through case studies, we assess whether existing guidelines fulfill system requirements, proposing improvements where necessary. In summary, our study focused on enhancing human-robot collaboration systems by incorporating improvements based on existing guidelines. We introduced functionalities to foster mutual understanding between humans and robots, aiming to mitigate psychological burdens and improve overall efficiency in the workplace.

Moving forward, our next step involves addressing requirements not covered in existing guidelines by developing the co-evolution guidebook [14]. This guidebook aims to complement existing guidelines by organizing requirements for lifecycle management of humans, machines, and environments. The co-evolution guidebook draws from existing guidelines while incorporating additional requirements. Currently, we are in the process of developing the guidebook and applying the co-evolution guidebook to the target system to assess its effectiveness and refine its application.

**Acknowledgments.** We express our heartfelt appreciation to Dr. Kenji Taguchi at UL Japan for his valuable insights and comments on the analysis conducted in this paper. His expertise and feedback have greatly enriched the quality and depth of our research. We are sincerely grateful for his contribution to our work.

# References

1. ISO/IEC: ISO/IEC TR 24368:2022 Information technology - Artificial intelligence - Overview of ethical and societal concerns, (2022)
2. IEC SEC 10: Ethics in Autonomous and Artificial Intelligence Applications, https://www.iec.ch/dyn/www/f?p=103:14:10280841609005::::FSP_ORG_ID,FSP_LANG_ID:22827,25, (Accessed April 2024)
3. High-Level Expert Group on Artificial Intelligence: Ethics guidelines for trustworthy AI, EUROPEAN COMMISSION, (2019)
4. ISO/IEC: ISO/IEC TR 24027:2021 Information technology - Artificial intelligence (AI) - Bias in AI systems and AI aided decision making, (2021)
5. ISO/IEC: ISO/IEC TR 24028:2020 Information technology - Artificial intelligence - Overview of trustworthiness in artificial intelligence, (2022)
6. ISO/IEC; ISO/IEC TR 24372:2021 Information technology - Artificial intelligence (AI) - Overview of computational approaches for AI systems, (2021)
7. ISO/IEC: ISO/IEC 38507:2022 Information technology - Governance of IT - Governance implications of the use of artificial intelligence by organizations, (2022)
8. ISO/IEC/IEEE: ISO/IEC/IEEE 24748-7000:2022 Systems and software engineering - Life cycle management - Part 7000: Standard model process for addressing ethical concerns during system design, (2022)
9. ISO/IEC: ISO/IEC TS 4213:2022 Information technology - Artificial intelligence - Assessment of machine learning classification performance, (2022)
10. S. Amershi, et. al: Guidelines for Human-AI Interaction, (2019)
11. S. Matsumoto, et. al: Shared Control in Human Robot Teaming: Toward Context-Aware Communication, arXiv:2023.10218v1 [cs.RO], (2022)
12. F. Semerato, et. al: Human-Robot Collaboration and Machine Learning: A Systematic Review of Recent Research, arXiv:2110.07448v1 [cs.RO], (2021)
13. E. Hollnagel: The Functional Resonance Analysis Method, A Handbook on how to use the FRAM, (2021)
14. Y. Matsubara, et. al: Toward Human-centered AI Framework: An Introduction to AI2X Co-evolution Project. SAFECOMP 2023, Toulouse, France. hal-04191518, (2023)
15. A FRAM Glossary: https://functionalresonance.com/a-fram-glossary.html (Accessed March 2024)
16. V. Voukalatou, et. al: Measuring objective and subjective well-being: dimensions and data source, International Journal of Data Science and Analytics, (2021)
17. ISO, ISO/DIS 25554 Ageing societies — Guidelines for promoting wellbeing in communities, (2023)